\documentclass[10pt,aps,pra,twocolumn,showpacs,superscriptaddress]{revtex4-2}
\usepackage[english]{babel} 
\usepackage[usenames, dvipsnames]{color} 
\usepackage{graphicx} 
\usepackage{bm} 
\usepackage{amsmath} 
\usepackage{amsthm} 
\usepackage{amssymb} 
\usepackage{times} 
\usepackage[pdftex,colorlinks=true,urlcolor= blue,linkcolor=Blue,citecolor=RedViolet]{hyperref}
\usepackage{multirow}
\usepackage{graphicx}
\usepackage{bm}
\usepackage{physics}
\usepackage{bbold}
\usepackage{comment}
\begin{document}

\providecommand{\moy}[1]{\langle #1 \rangle}
\providecommand{\bra}[1]{\langle #1 |}
\providecommand{\ket}[1]{| #1 \rangle}
\providecommand{\braket}[2]{\langle #1 | #2 \rangle}
\providecommand{\ketbra}[2]{|  #1\rangle \langle #2 |}
\newcommand{\aspasc}{\textquotedblleft}
\newcommand{\aspasd}{\textquotedblright\ }
\newcommand{\blue}[1]{\textcolor{blue}{#1}}
\newcommand{\red}[1]{\textcolor{red}{#1}}

\title{Ergotropic and passive contributions on the phase-space information geometry of Gaussian states}
 \author{Ivan \surname{Medina}}
 \email{ivan.medina@ifsc.usp.br}
 \affiliation{S\~{a}o Carlos Institute of Physics, University of S\~{a}o Paulo, S\~{a}o Carlos 13566-590, S\~{a}o Paulo, Brazil}
 \author{Camila Raupp}
 \affiliation{S\~{a}o Carlos Institute of Physics, University of S\~{a}o Paulo, S\~{a}o Carlos 13566-590, S\~{a}o Paulo, Brazil}
  \author{Pedro B. Melo}
  \affiliation{Department of Physics, PUC-Rio, Rio de Janeiro 22451-900 RJ, Brazil}
  \affiliation{Universit\`a degli Studi di Palermo, Dipartimento di Fisica e Chimica - Emilio Segr\`e, via Archirafi 36, I-90123 Palermo, Italy}
 \author{Diogo \surname{O. Soares-Pinto}}
 \affiliation{S\~{a}o Carlos Institute of Physics, University of S\~{a}o Paulo, S\~{a}o Carlos 13566-590, S\~{a}o Paulo, Brazil}

\begin{abstract}
We establish a direct connection between quantum thermodynamics and information geometry by introducing an ergotropic decomposition of the Wigner-Fisher information for Gaussian quantum states. We derive an analytical expression for the Wigner-Fisher information in terms of the phase-space covariance matrix and mean vector, and show that it naturally separates into passive and ergotropic contributions. The passive term is shown to be entirely determined by the Wigner entropy rate of the associated passive state. The ergotropic contribution, in turn, quantifies how displacement and squeezing resources modify the statistical velocity and length of the system trajectory in phase space. As an application, we analyze the recently proposed ergotropic Mpemba effect and demonstrate that it can be traced to the anomalous geometric evolution of the passive state associated with squeezed thermal states. Our results reveal how the extractable work stored in a quantum state constrains its information-geometric structure, establishing a framework that links ergotropy, entropy production rate, and statistical geometry in continuous-variable quantum systems.
\end{abstract}
\maketitle





\begin{figure}[t]
    \centering
    \includegraphics[width=\linewidth]{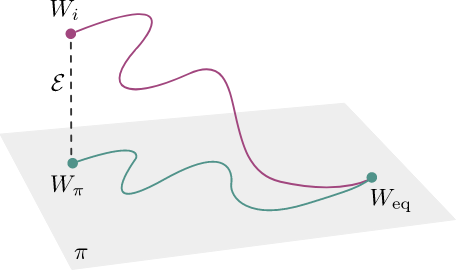}
    \caption{The trajectory of a Gaussian quantum state on the Wigner function statistical manifold is characterized by the Wigner- Fisher information, which can be separated into passive and ergotropic contributions. The passive component is associated with the evolution of the corresponding passive state along the thermal-state manifold represented by the surface $\pi$, while the ergotropic component captures the geometric effect of energy stored through displacement and squeezing operations.}
    \label{fig1}
\end{figure} 

The connection between  thermodynamics and information theory has provided important insights on seminal problems such as the interpretation of the Second Law and the Maxwell demon paradox \cite {Jaynes1957I,Jaynes1957II,leff1990maxwell, Maruyama2009}. More recently,  thermodynamics has been explored using information geometry techniques \cite{amari2000methods}, which relate the distinguishability of probability distributions to thermodynamic quantities \cite{Crooks2007,Feng2009,Sivak2012_2,Ito2018,Ito2020_2,Hasegawa2021,Dechant2022,Bettmann2025}. This link relies on the geometric interpretation of the Fisher information \cite{Braunstein1994}. Classically,  Chentsov's theorem \cite{cencov2000statistical, Dowty2018} establishes the Fisher information as the unique Riemannian metric contractive under stochastic (noisy) operations, allowing one to treat the evolution of a system as a trajectory described by  probability distributions on a statistical manifold \cite{Bengtsson2017}. In the quantum realm, this analysis is more delicate - there is a family of Fisher information which are contractive under CPTP maps and can be used as metrics on the space of quantum states \cite{morozova1991markov,petz1996monotone,petz1996riemannian}. This informational perspective yielded important contributions on the limitations of physical processes, such as 
the development of generalized uncertainty relations \cite{Nicholson2018,Vu2019,Vu2019_2,Nicholson2020,Saito2023} and  geometric speed limits \cite{delCampo2013,Taddei2013,Pires2016,Shiraishi2018,Okuyama2018,Ito2020,Hasegawa2020,Pintos2022,Dong2025,Melo2025,melo2026}. It also has led to deeper understanding of complex thermodynamic processes such as dissipative phase transitions \cite{bettmann2025thermodynamic} and transitions between non-equilibrium steady states \cite{lacerda2025information}.



From the perspective of quantum thermodynamics, a central goal is to identify the operational value of nonequilibrium states. A key quantity in this context is the ergotropy \cite{Allahverdyan2004}, defined as the maximum amount of work that can be extracted from a quantum system through cyclic unitary operations. Ergotropy has proven particularly useful for quantum batteries \cite{CampaioliRev2024}, where it represents the amount of charge a system can store. On a more  fundamental level, an ergotropy-based formulation of thermodynamics was recently proposed in Ref. \cite{Choquehuanca2025}, establishing a connection between heat and the evolution of the passive state, while also providing a thermodynamical way for quantifying non-Markovianity. Despite these advances, the study of ergotropy in open quantum systems remains in its early stages, and its connections to other quantum resources are still poorly understood. Although the relationship between ergotropy and relative entropies is established for thermal reference states \cite{Plastina2020,Sone2021,Medina2025}, to the best of our knowledge, the connection between ergotropy and information geomerty remains unexplored.

In this Letter, we provide a direct connection between ergotropy and information geometry. For this, we study the evolution of a single bosonic mode, restricting our model to the class of Gaussian states and gaussianity preserving operations.  This class of continua variable states is particularly interesting as phase-space tomography techniques are experimentally available on several platforms. Also, the simplicity of Gaussian processes, which depends only on a mean vector and a covariance matrix, has provided insightful analytical results for thermodynamic protocols involving energy extraction through ergotropy \cite{Brown_2016,Singh2019,Medina2025,Rodriguez2025} and also for quantum metrology \cite{Slater2006,Fabre2012,Pinel2013,Zhang2013,Deffner_2017,Vershynina2017,Tang_2025,RevFadel_2025}. By constructing our problem in the phase space representation, we first derive the basis independent Wigner-Fisher information (WFI), which is directly related to the Wigner function of a quantum state. We focus on the WFI parameterized by the single parameter - time, and introduce geometric quantities such as the statistical length and velocity. As depicted in Fig. \ref{fig1}, our main result shows that the WFI can be decomposed into passive and ergotropic contributions, revealing the role of extractable work in the trajectories traced on the statistical manifold. As an application, we study the recently proposed ergotropic Mpemba effect from an information-geometric perspective, showing that its origin can be traced to the anomalous evolution of the associated passive state in thermal state manifold.

\paragraph*{Wigner-Fisher information for Gaussian states -} We start by defining the Wigner-Fisher information. For this, we consider a single bosonic mode described by the Hamiltonian ($\hbar=1$) $\hat{H}=\omega(\hat{a}^\dagger\hat{a}+1/2)$, where $\hat{a}$ ($\hat{a}^\dagger$) are the annihilation (creation) operator satisfying $[\hat{a},\hat{a}^\dagger]=1$. Moreover, we assume that our system is described by Gaussian states, which are fully characterized by a mean vector $\vec{v}_\theta=(\moy{\hat{a},\hat{a}^\dagger})$ and the covariance matrix $\Theta_\theta$ with elements $[\Theta_\theta]_{ij}=\frac{1}{2}\moy{u_i u_j^\dagger+u_j^\dagger u_i}-\moy{u_i}\moy{u_j^\dagger}$, with $\vec{u}=(\hat{a},\hat{a}^\dagger)$. Here we assume that our Gaussian state is parameterized by a single parameter $\theta$. Any Gaussian state can be represented in the phase-space by the positive Wigner function
\begin{align}
    W_\theta(\vec{\alpha})=\frac{1}{\pi \sqrt{\det\Theta_\theta}}\exp\left[-\frac{1}{2}(\vec{\alpha}-\vec{v}_\theta)^\dagger\Theta^{-1}_\theta(\vec{\alpha}-\vec{v}_\theta)\right],\nonumber
\end{align}
where $\vec{\alpha}=(\alpha,\alpha^*)$. Since $W_\theta(\vec{\alpha})>0$, we can define a Wigner relative entropy
\begin{align}
    K[W_1||W_2]=\int d^2\alpha W_1(\vec{\alpha})\ln\frac{W_1(\vec{\alpha})}{W_2(\vec{\alpha})},
\end{align}
which is the  Kullback-Leibler divergence between any two probability distributions represented by positive Wigner functions $W_{1}(\vec{\alpha})$ and $W_{2}(\vec{\alpha})$. 

Considering the Wigner function computed for a small deviation $\theta+d\theta$, we can expand the Wigner relative entropy between $W_\theta(\vec{\alpha})$ and $W_{\theta+d\theta}(\vec{\alpha})$ up to second order in $d\theta$ as $K[W_\theta||W_{\theta+d\theta}]=\frac{1}{2}I_{\rm W}(\theta)d\theta^2+O(d\theta^3)$, where we define the WFI as $I_{\rm W}(\theta)=\int d^2\alpha W_{\theta}(\vec{\alpha})[\frac{d}{d\theta}\ln W_{\theta}(\vec{\alpha})]^{2}$. The WFI can be analytically computed in therms of $\vec{v}_\theta$ and $\Theta_\theta$ [See Appendix \ref{AppA}], 
\begin{equation}
I_{\rm W}(\theta)=\dot{\vec{v}}_{\theta}^{\dagger}\Theta_{\theta}^{-1}\dot{\vec{v}}_{\theta}+\text{Tr}\{(\Theta_{\theta}^{-1}\dot{\Theta}_{\theta})^{2}\},
\label{WFI}
\end{equation}
and due to its geometric interpretation and connection to the Wigner relative entropy, it constitutes this work main tool to study thermodynamics geometry in the phase-space of Gaussian states. The Wigner distribution is a basis independent quantity that contains all the information about the system state. Moreover, as $W_\theta(\vec{\alpha})$ behaves essentially as a classical probability distribution, the Chentsov's theorem guarantees that the WFI is the only Riemannian metric on the statistical manifold \cite{cencov2000statistical, Dowty2018}.  Thus it happens that the WFI is identical to the Symmetric Logarithm Derivative (SLD) quantum Fisher information when working with the state in the Hilbert space \cite{Zhang2013,Monras2013,Nielsen2023,Sorelli2024}. This means that it gives the optimal bound in the estimation of the single parameter $\theta$, i.e., it obeys the Cramér–Rao bound ${\rm Var(\theta)}\geq [M I_{\rm W}(\theta)]^{-1}$, with $M$ being the number of independent measurements.

\paragraph*{Temporal Wigner-Fisher information -}  By considering time as a parameter ($\theta\rightarrow t$) and taking into account that the WFI is a metric, the square of the line element, $ds$, between two Wigner functions infinitesimally displaced from one another on the manifold is defined as 
\begin{align}
    ds^2=\frac{1}{2}I_{\rm W}(t)dt^2.
    \label{dssquare}
\end{align}
Then, we can define a path length over a trajectory $\gamma$ as
\begin{align}
\mathcal{L}_{\rm W}(t) =\int_\gamma ds=\int_0^t \frac{ds}{d\tau}d\tau =\frac{1}{\sqrt{2}}\int_{0}^{t} d\tau \sqrt{I_{\rm W}(\tau)},    \label{flenght}
\end{align}
which represents the statistical distance in the phase-space distributions. We remark that  $V_{\rm W}(t) = \frac{ds}{dt}=\frac{1}{\sqrt{2}}\sqrt{I_{W}(t)}$ can be interpreted as a statistical velocity, i.e., it express the rate at which the the distribution $W_t(\vec{\alpha})$ changes in the statistical manifold. It is usually convenient to define a degree of completion
\begin{align}
\eta_\tau(t)=\frac{\mathcal{L}_{\rm W}(t)}{\mathcal{L}_{\rm W}(\tau)},\label{doc}
\end{align}
which represents the amount of the length traced in the statistical manifold until a time $t$ with respect to the total length for a fixed final time $\tau$.

In this work we consider our system weakly coupled to a thermal reservoir. Under this condition it obeys the GKLS master equation
\begin{align}
    \dot{\hat{\rho}}_t=-i[\hat{H},\hat{\rho}_t]+\Gamma(\bar{n}_{\rm eq}+1)\mathcal{D}_\rho[\hat{a}]+\Gamma\bar{n}_{\rm eq}\mathcal{D}_\rho[\hat{a}^\dagger],\label{meq}
\end{align}
where $\Gamma$ is a dissipation rate, $\bar{n}_{\rm eq}=(e^{\omega \beta_{\rm eq} } - 1)^{-1}$ is the thermal occupation number related to the thermal fixed point of the dynamics and $\mathcal{D}_\rho[\hat{L}]=\hat{L}\hat{\rho} \hat{L}^\dagger-\frac{1}{2}\{\hat{L}^\dagger \hat{L}, \hat{\rho}\}$ is the dissipator. From this master equation, the mean vector evolves as $\dot{\vec{v}}_t=\Lambda\vec{v}_t$ and the covariance matrix obeys the Lyapunov Equation $\dot{\Theta}_t=\Lambda\Theta_t+\Theta_t\Lambda^\dagger+F$, with
\begin{align}
\Lambda=-\frac{1}{2}
\begin{bmatrix}
\Gamma+2i\omega & 0 \\
0 & \Gamma-2i\omega\\
\end{bmatrix}, \ \ F=\Gamma\left(\bar{n}_{\rm eq}+\frac{1}{2}\right)\mathbb{I},
\end{align}
where $\mathbb{I}$ is the dim-2 identity matrix. This dynamics satisfies the detailed balance condition, and allows a clear thermodynamic connection with the WFI.

\paragraph*{Ergotropic and passive contributions to the WFI - } We now discuss our main result, which is the connection of the WFI with the ergotropy. The ergotropy has a clear thermodynamic meaning, as it represents the maximum amount of work that can be extracted from the system by means of cyclic unitary operations \cite{Allahverdyan2004}. In Ref. \cite{Medina2025}, it is shown that for Gaussian states ergotropy is given by $\mathcal{E}=\omega(\bar{n}_\pi+1/2)K[W||W_\pi]$, where  $W_\pi(\vec{\alpha})$ is the phase space representation of the thermal passive state $\rho_{\pi}$ characterized by the thermal ocupation number $\bar{n}_\pi$. Naturally, the dissipation caused by the interaction of the system with the environment induces an ergotropic loss rate, which can be expressed as [See Appendix \ref{AppC}]
\begin{align}
        \dot{\mathcal{E}}(t)=-E_{\rm eq}[\Pi_{\rm W}(t)-\Pi_{ {\pi}}(t)],\label{ergrate}
\end{align}
where $E_{\rm eq}=\omega(\bar{n}_{\rm eq}+1/2)$ is the equilibrium state mean energy and $\Pi_{\rm W}(t)=-\frac{d}{dt}K[W||W_{\rm eq}]$ is the Wigner entropy production rate as defined in Ref. \cite{Santos2017}. The term $\Pi_{{\pi}}(t)$ is the Wigner entropy production rate of the passive state. The result in Eq. \eqref{ergrate} shows that the rate of energy that can be unitarily extracted from Gaussian states is proportional to the difference between the entropy produced by the state and the entropy produced by its associated passive state. This is an important result, since it means that for an open quantum system, the evolution and amount of entropy produced by the passive state play a crucial role in characterizing the ergotropy loss rate. As we shall see,  WFI captures the entropic contributions of the passive and ergotropic parts of the state, allowing a clear energetic and geometric interpretation for thermalization of Gaussian state.

A general Gaussian state can be parameterized as $\hat{\rho}=\hat{D}(\mu)\hat{S}(z)\hat{\rho}_{\pi}\hat{S}^\dagger(z)\hat{D}^\dagger(\mu)$, where $D(\mu)=e^{\mu\hat{a}^\dagger-\mu^*\hat{a}}$ and $\hat{S}(z)=\exp\left[\frac{1}{2}(z\hat{a}^{\dagger 2}-z^*\hat{a}^2)\right]$ are the displacement and squeezing operators. The complex parameters $\mu=|\mu|e^{i\frac{\phi_{\rm d}}{2}}$ and $z=re^{i\phi_s}$  ($r\equiv|z|$) are defined by   the mean vector modulus, $|\mu|$, and squeezing parameter $r$; $\phi_{\rm d}$ and $\phi_{\rm s}$ are real phases. The main contribution of this letter is to show that for gaussian states the WFI can be written as $I_{\rm W}(t)=I_{\pi}(t)+I_{\mathcal{E}}(t)$, where $I_{\pi}(t)$ accounts for the passive state (thermal) contribution and $I_{\mathcal{E}}(t) = I_{\rm d}(t)+I_{\rm s}(t)+I_{\rm ds}(t)$ quantifies how the energy invested by unitary Gaussian operations modifies the WFI. Using the Lyapunov equation and the expression in Eq. \eqref{WFI} for the WFI we can write  [see Appendix \ref{AppB}] 
\begin{align}
 &I_{\pi}(t)=\frac{1}{2}\left[\frac{d}{dt}\ln(\det\Theta_t)\right]^2=2\dot{S}_{\rm W}^2(t),\\
 &I_{\rm d}(t) = \frac{|\dot{\vec{v}}_t|^2}{\sqrt{\det \Theta_t}},\label{Id}\\
 &I_{\rm s}(t)={\rm Tr}\{(\Theta^{-1}_t\dot{\Theta}_t)^2\}-\frac{1}{2}({\rm Tr}\{\Theta^{-1}_t\dot{\Theta}_t\})^2,\label{Is}\\
 &I_{\rm ds}(t) = \frac{\left[({\rm Tr\{\Theta_t\}}-\sqrt{\det\Theta_t})|\dot{\vec{v}}_{t}|^2-\dot{\vec{v}}_{t}^{\dagger}\Theta_{t}\dot{\vec{v}}_{t}\right]}{\det\Theta_t}.\label{Ids}
\end{align}
where $ S_{\rm W}(t)=-\int d^2\alpha W_t(\vec{\alpha})\ln W_t(\vec{\alpha})$ is the Wigner entropy. Let us first focus on the passive contribution. As for any Gaussian state we have an associated passive state satisfying $\det \Theta=\det \Theta_\pi$ for all times, the passive  contribution, $I_{\pi}(t)$, accounts for the entropy rate contribution to the total WFI. More than that, as we can associate a physical state to the passive state, it means that the passive contribution $I_{\pi}(t)$ represents the total WFI for the thermal state with thermal distribution determined from  $\sqrt{\det \Theta_\pi}=\bar{n}_\pi(t)+1/2$. Note that this is not true, in general,  for the ergotropic contribution, i.e., we cannot associate a physical state to the contribution $I_\mathcal{E}(t)$ alone. This is possible only for a very special case that will be discussed later. As $I_\pi(t)$ represents a real WFI, we can define, as in Eq. \eqref{flenght}, the statistical length
\begin{align}
   \mathcal{L}_{\rm \pi}(t) =\frac{1}{\sqrt{2}}\int_{0}^{t} d\tau \sqrt{I_{\pi}(\tau)}=\int_{0}^{t} d\tau |\dot{S}_{\rm W}(t)|,  
\end{align}
which has a clear interpretation: it represents the path length that the passive state follows in the manifold composed only by thermal states. Moreover, the mean energy of the passive state is given by $E_{\pi}(t)=\omega\sqrt{\det \Theta_t}$. Therefore, the Wigner entropy rate is equal to the instantaneous relative energy rate,  $\dot{S}_{\rm W}(t)=\dot{E}_\pi(t)/E_\pi(t)$.

To clearly understand the role of the passive contribution, consider that the system's initial state is a passive state, i.e., $\hat{\rho}=\hat{\rho}_\pi$. As thermal states have a null mean vector and a covariance matrix that is proportional to the identity operator, it immediately reads that $I_{\rm ds}(t) = I_{\rm d}(t)=I_{\rm s}(t)=0$, as expected. Then, the dynamics of the system is characterized by $\Theta_t=[\bar{n}_{\pi}(t)+\frac{1}{2}]\mathbb{I}$ where the mean occupation number $\bar{n}_\pi(t)=[\bar{n}_\pi(0)-\bar{n}_{\rm eq}]e^{-\Gamma t}+\bar{n}_{\rm eq}$ defines an effective time dependent temperature.  
For this case, $S_{\rm W}(t)$ is a monotonic quantity and the statistical path length can be easily obtained 
\begin{align}
 \mathcal{L_\pi}(t)=|S_{\rm W}(t)-S_{\rm W}(0)|\label{thpath}. 
\end{align}
This means that for an initial thermal state the path length is given by the absolute value of the entropy variation of the system.

Now, for the ergotropic terms in Eqs. \eqref{Id}-\eqref{Ids},  $I_{\rm d}(t)$ and $I_{\rm s}(t)$ represent the contributions to the WFI coming from the displacement and squeezing operations separately while $I_{\rm ds}(t)$ represents a crossing contribution between displacement and squeezing so that $I_{\rm ds}(t)=0$ when $\mu=0$ or $r=0$. To establish a clear thermodynamic connection, in what follows we study each contribution separately.

When the system is initially in a displaced thermal state, $\hat{\rho}=\hat{D}(\mu)\hat{\rho}_{\pi}\hat{D}^\dagger(\mu)$, ergotropy is stored in the phase-space mean vector as $\mathcal{E_{\rm d}}(t)=\omega|\mu|^2e^{-\Gamma t}$. From Eq. \eqref{Id}, we can compute the displacement contibution to the WFI
\begin{equation}
I_{\rm  d}(t)=2\left[\omega^2+\frac{\Gamma^2}{4}\right]R_{\rm d}(t),
\end{equation}
where $R_{\rm d}(t)=\mathcal{E}_{d}(t)/E_{\pi}(t)$ quantifies the ratio of available ergotropy to the passive thermal energy. This result tells us that the displacement contribution is proportional to the amount of energy stored in the system state. Interestingly, as the covariance matrix is not affected by the displacement operation, the associated passive state follows exactly the same statistical path as if the initial state were a thermal state [see Eq. \eqref{thpath}]. As we shall see, this is not true for squeezed states, since the squeezing operation affects the covariance matrix and, by consequence, the effective temperature evolution of the passive state.

A particular and interesting case of displaced thermal state is when the passive state is the equilibrium state itself, i.e., the system is initially prepared in $\hat{\rho}=\hat{D}(\mu)\hat{\rho}_{\rm eq}\hat{D}^\dagger(\mu)$. As the displacement does not affect the covariance matrix, we have that $\Theta_t$ is constant leading to $\dot{S}_{\rm W}(t)=0$ and consequently $I_\pi(t)=0$. For this particular case, we have that WFI is purely ergotropic, i.e., $I_{\rm W}(t)=I_{\rm d}(t)$. This allows us to analytically compute the ergotropic path length, which is given by
\begin{align}
     \mathcal{L}_{\rm d}(t)& =2\sqrt{\frac{|\mu|(\Gamma^2+\omega^2)}{\Gamma^2(2+4\bar{n}_{\rm eq})}}(1-e^{-\frac{\Gamma t}{2}}).
\end{align}



For a system initially in a squeezed thermal state, $\hat{\rho}=\hat{S}(z)\hat{\rho}_{\rm \pi}\hat{S}^\dagger(z)$, ergotopry is stored in the covariance matrix only, since the squeezing operation does not affect the mean vector of a thermal state. From Eq. \eqref{Is}, we obtain
\begin{equation}
I_{\rm s}(t)=\frac{2\dot{R}_s^{2}(t)}{R_s(t)[2+R_s(t)]}+8\omega^2R_s(t)[2+R_s(t)],
\end{equation}
where $R_{\rm s}(t)=\mathcal{E}_{s}(t)/E_{\pi}(t)$ with ergotropy stored by squeezing being $\mathcal{E}_{s}(t)=E_{\pi}(t)[\cos(2 r_t)-1]$. Fundamentally distinct from the displacement operation, squeezing affects the thermal occupation number of the passive state, which acquires a non trivial dependence on time. From the Lyapunov equation, we have that the passive state thermal occupation number is given by
\begin{align}
&\bar{n}_\pi(t)+\frac{1}{2}=\sqrt{\left[\bar{n}_{\rm th}(t)+\frac{1}{2}\right]^2+\bar{n}_{\rm s}e^{-2\Gamma t}(e^{\Gamma t}-1)}, \label{nsqueeze}
\end{align}
where $\bar{n}_{\rm th}(t)=[\bar{n}_\pi(0)-\bar{n}_{\rm eq}]e^{-\Gamma t}+\bar{n}_{\rm eq}$ is the solution for the mean thermal occupation when no squeezing is present and $\bar{n}_s=(2\bar{n}_\pi(0)+1)(2\bar{n}_{\rm eq}+1)\sinh^2(r)$ is the squeezing effect. The consequence of this results is that the squeezing affects the path that the passive state follows on the thermal state manifold. In this case, the entropy is not a monotonic function of time, then a trivial solution such as in Eq. \eqref{thpath} is not valid in general. We remark that differently for what happens for an initial displaced thermal state when $\rho_\pi=\rho_{\rm eq}$, as the squeezing affects $\bar{n}_\pi(t)$, we now have that $S_{\rm W}(t)$ is not constant, thus $I_{\pi}(t)\neq0$ in general.




Finally, for a general squeezed-displaced thermal state, a crossing term involving ergotropic contributions from displacement and squeezing arises. From Eq. \eqref{Ids}, it reads
\begin{equation}
     I_{\rm ds}(t)=I_{\rm d}(t)\left[R_{{\rm s}}(t)+\sqrt{R_{{\rm s}}(t)[R_{{\rm s}}(t)+2]}\cos(\Delta-\varphi)\right],
\end{equation}
where $\Delta = \phi_{\rm d} - \phi_{\rm s}$ and $ \varphi=\arctan\left[\frac{\Gamma}{4\omega}-\frac{\omega}{\Gamma}\right]$. As we see, although the phases $\phi_{\rm d}$ and $\phi_{\rm s}$ do not contribute for the ergotropy or energy of Gaussian states, they do contribute for the WFI. This means essentially that there is a class of Gaussian states energetically equivalent, but describing different paths on the Wigner function manifold. Moreover, the WFI does not involve a simply sum of terms coming from displacement and squeezing separately, it also involves a non trivial crossing term.

\begin{figure}[t]
    \centering
    \includegraphics[width=\linewidth]{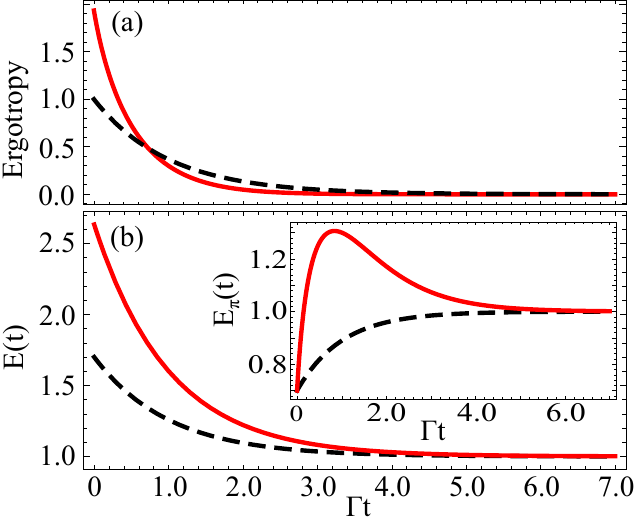}
    \caption{(a) Example of the ergotropic Mpemba effect. Considering two different initial states, $\rho_{\rm s}$ and $\rho_{\rm d}$, we see that although $\rho_{\rm s}$ has initially more ergotropy than $\rho_{\rm d}$, a squeezing thermal state loses charge exponentially faster than the displaced thermal state, giving rise to the effect. (b) Mean energy of $\rho_{\rm s}$ and $\rho_{\rm d}$ and their associated passive mean energy (inset). We see that the passive energy associated to $\rho_{\rm s}$ is non monotonic in time. For all plots, quantities for $\rho_{\rm s}$ are shown in red-solid line and quantities for $\rho_{\rm d}$ are shown in black-dashed lines. The parameters we use are $\omega=1.0$, $\Gamma=0.1$, $r=1.0$, $\mu=1.0$, $\bar{n}_{\rm eq}=0.5$ and $\bar{n}_\pi(0)=0.2$.  }
    \label{fig2}
\end{figure}

\begin{figure*}[t]
    \centering
    \includegraphics[width=\linewidth]{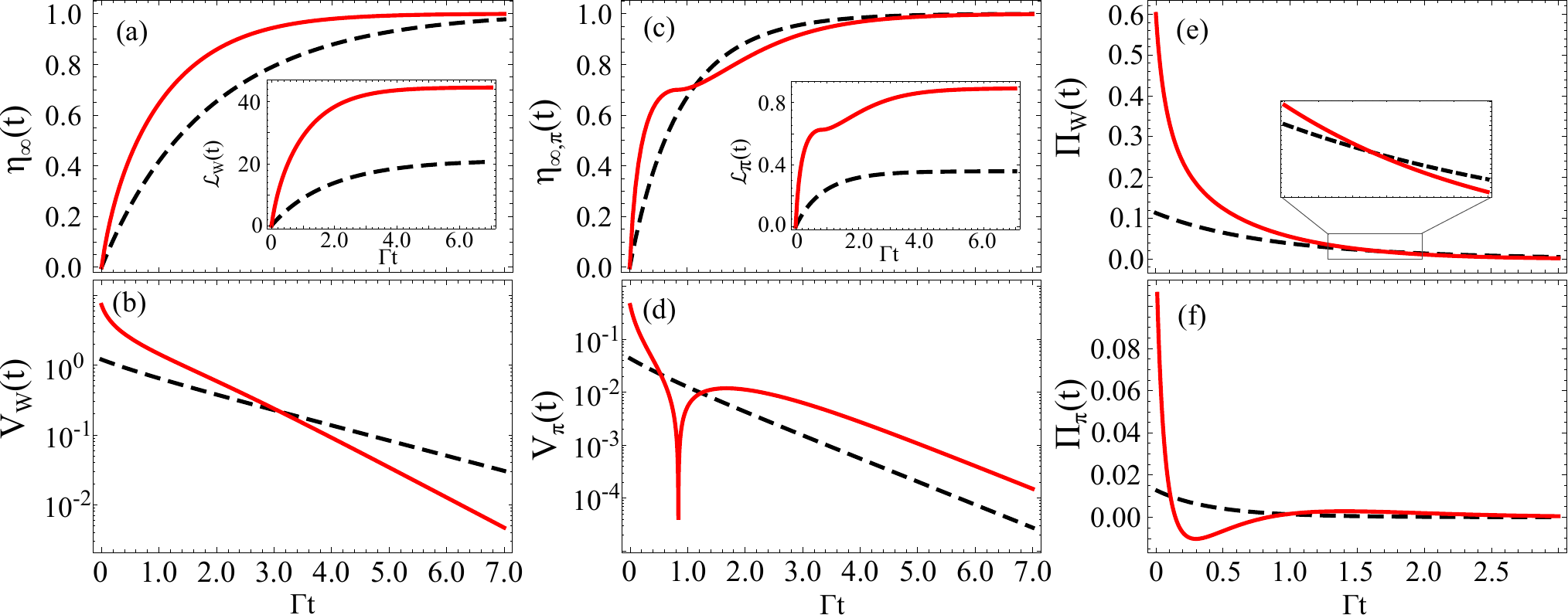}
    \caption{Degree of completion and statistical length as a function of time. Plots (a) and (b), respectively, show the quantities for $\rho_{\rm d}$ and $\rho_{\rm s}$ while the plots (c) and (d) are related to the passive states. We see that the degree of completion and statisical velocities for the passive state $\rho_\pi^{\rm s}$ are not monotonic. This behavior happens because $\rho_\pi^{\rm s}(t)$ is not a solution of the master Eq. \eqref{meq}.  The plots (e) and (f) show the entropy production rates $\Pi_{\rm W}(t)$ and $\Pi_\pi(t)$. The negative values reached by $\Pi_\pi(t)$ suggests that the passive state $\rho_\pi^{\rm s}(t)$ follows a non-Markovian dynamics.  We use the same parameters as in Fig. \ref{fig2}. }
    \label{fig3}
\end{figure*}

\paragraph*{Example: ergotropic Mpemba effect - } The Mpemba effect is an anomalous relaxation process commonly known as hot water freeze faster than warm water \cite{Mpemba1969,Kell69}. This effect has recently been extended and extensively explored in the quantum domain due to its fundamental implications and possible technological applications \cite{Moroder2024,Medina2025,Ares2025,Summer2026,Westhoff2025}. In particular, in Ref. \cite{Medina2025}, the ergotropic Mpemba effect was introduced. In the context of quantum batteries, ergotropy represents the amount of charge in a quantum system. It was found that systems charged initially with more ergotropy stored as squeezing, discharge faster than systems initially with a lower charge in form of displacement. In Fig. \ref{fig2} we show an example of the ergotropic Mpemba effect. We consider two different initial states, a displaced thermal state, $\rho_{\rm d}$, and a squeezed thermal state $\rho_{\rm s}$. We also choose $\mu=1.0$ and $r=1.0$ such that $\mathcal{E}_{\rm s}(0)>\mathcal{E}_{\rm d}(0)$. In all plots the red-continuum line represents the quantities computed from $\rho_{\rm s}$ and the black-dashed lines $\rho_{\rm d}$. We also use indices d and s to distinguish quantities computed from these two states.  From Fig. \ref{fig2} (a) we see the ergotropic Mpemba effect occurring: although $\rho_{\rm s}$ has initially more ergotropy, this state loose charge exponentially faster than $\rho_{\rm d}$ and a Mpemba crossing happens around $\Gamma t=1.0$.  As discussed in Ref. \cite{Medina2025}, this effect is a consequence of the evolution of the passive states. Although the energy of $\rho_{\rm d}$ and $\rho_{\rm s}$ dissipates at the same rate, the energy associated with their passive states are distinctly different, as can be seen from Fig. \ref{fig2} (b) and inset. Due to the effect of squeezing on the thermal occupation of its passive state, $\rho_\pi^{\rm s}$, the passive energy $E_{\pi}^{\rm s}(t)$ has a non-monotonic behavior towards the equilibrium and satisfies $E_{\pi}^{\rm s}(t)\geq E_{\pi}^{\rm d}(t)$. As the ergotropy is given by the difference $E(t)-E_\pi(t)$, the ergotropy of $\rho_{\rm s}$ will always have a higher contribution coming from its passive state, making it relaxes faster than the ergotropy for $\rho_{\rm d}$.

We now explore this effect under the light of the WFI.  Our splitting into ergotropic and passive parts allows us to study not only the path length and velocity of the total states, but also the statistical quantities of their associated passive states along the thermal states manifold. We recall that $I_\pi(t)$ depends essentially on the instantaneous relative passive energy rate $\dot{E}_{\pi}(t)/E_\pi(t)$, and this will help us to understand the anomalous behavior of the passive state associated to $\rho_{\rm s}$. In Fig. \ref{fig3} (a) and its inset, it is shown that although $\rho_{\rm s}$ traces a longer path on the statistical manifold, it reaches the equilibrium faster than $\rho_{\rm d}$. This is evidenced by $\eta_\infty(t)$, which represents how close the system is to the equilibrium state [see Eq. \eqref{doc}]. This is explained by $V_{\rm W}^{\rm s}(t)$, which decays exponentially faster than $V_{\rm W}^{\rm d}(t)$, as we can see from Fig. \ref{fig3} (b).

Let us now look at the contributions from the passive state. In Fig. \ref{fig3} (c), we see that due to the effect of the squeezing operation on the thermal occupation [see Eq. \eqref{nsqueeze}] both  $\mathcal{L}_{\rm \pi}^{\rm s}(t)$ and $\eta_{\infty,\pi}^{\rm s}(t)$ do not have a monotonic behavior. In the beginning of the dynamics we observe that $\rho_{\rm \pi}^{\rm s}(t)$ approaches the equilibrium faster than $\rho_{\rm \pi}^{\rm d}(t)$. However, for $\Gamma t\approx 0.6$, $\eta_{\infty,\pi}^{\rm s}(t)$ changes its behavior crossing $\eta_{\infty,\pi}^{\rm d}(t)$ around $\Gamma t=1.0$. Then, $\rho_{\rm \pi}^{\rm s}(t)$ ultimately takes a longer time to reach the equilibrium state. This behavior is better understood if we look at $V_{\pi}^{\rm s}(t)$.  We see that $V_{\pi}^{\rm s}(t)$ decreases exponentially faster than $V_{\pi}^{\rm d}(t)$ until $\Gamma t\approx1.0$, where $V_{\pi}^{\rm s}(t)\approx0$. After that, the passive state ``accelerates" as $V_{\pi}^{\rm s}(t)$ increases until $\Gamma t\approx1.5$, where it finally backs to exponentially decreases at the same rate as $V_{\pi}^{\rm d}(t)$. This acceleration means that for this period of time the passive state is pushed far from the equilibrium state. This corroborates with the  behavior of $E_\pi^{\rm}(t)$ [seen inset of Fig. \ref{fig2} (b)] which increases far from the equilibrium energy (for this example, $E_{\rm eq}=1.0$), before it relaxes towards it. This anomalous behavior follows from the fact that $\rho_{\pi}^{s}(t)$ is not a solution of the Markovian Eq. \eqref{meq}. In fact,  $\rho_\pi^{\rm s}(t)$ present some dynamical signatures that are characteristic from non-Markovian dynamics. In Fig. \ref{fig3} (e) we see that as expected from a Markovian dynamics, both the entropy production rates $\Pi_{\rm W}^{\rm s}(t)$ and $\Pi_{\rm W}^{\rm d}(t)$ decay monotonically.  However, in Fig. \ref{fig3} (f) we see that $\Pi_{\pi}^{\rm s}(t)$ has a non-monotonic behavior and also reaches negative values, which is characteristic from non-Markovian dynamics. The above discussion and results suggest that along the dynamics part of squeezing is converted into thermal energy for the passive state, simulating a non-Markovian effect and giving rise to the ergotropic Mpemba effect.

\paragraph*{Conclusions -} In this Letter, we established a direct connection between ergotropy and information geometry by deriving an exact decomposition of the Wigner-Fisher information into passive and ergotropic contributions for Gaussian quantum states. This decomposition provides a thermodynamic interpretation for the statistical length and velocity associated with thermalization processes, identifying the distinct roles played by entropy production and extractable work in the trajectories on the manifold of Gaussian states. As an application, we showed that the ergotropic Mpemba effect can be understood in geometric terms through the anomalous evolution of the associated passive state. Our results place ergotropy within the framework of information geometry and open new avenues for investigating thermodynamic bounds, nonequilibrium phenomena, and resource-theoretic aspects of continuous-variable quantum systems.




\paragraph*{Acknowledgments} - The authors thank Steve Campbell, Mauro Paternostro, Lucas C\'eleri and Fernando Nic\'acio for fruitful discussions. I.M. acknowledges financial support from the São Paulo Research Foundation – FAPESP (Grant No. 2022/08786-2 and No. 2023/14488-7). C.R.L acknowledges the support of Brazilian agencies CAPES - Finance Code 001.  P.B.M. acknowledges the support of Brazilian agencies CAPES - Finance Code 001, and CNPq, grant No. 140264/2026-4. D.O.S.P. acknowledges support from the Brazilian
funding agency CNPq (Grants No. 304891/2022-3).

\providecommand{\newblock}{}
\bibliography{refs.bib}

\appendix
\newpage
\onecolumngrid

\section{Wigner-Fisher information for Gaussian States}
\label{AppA}

In this appendix, we derive the Wigner-Fisher information expression for a single-bosonic mode described by a Gaussian state. Any single mode Gaussian state can be expressed in the phase space representation by a Wigner function as
\begin{align}
W_\theta(\vec{\alpha})=\frac{1}{\pi \sqrt{\det\Theta_\theta}}\exp\left[-\frac{1}{2}(\vec{\alpha}-\vec{v}_\theta)^\dagger\Theta^{-1}_\theta(\vec{\alpha}-\vec{v}_\theta)\right],\label{gaussian}
\end{align}
where $\vec{\alpha}=(\alpha,\alpha^*)$, $\vec{v}_\theta=(\langle a\rangle,\langle a^\dagger\rangle)$ is the mean vector and
\begin{align}
\Theta_\theta=
\begin{bmatrix}
\frac{1}{2}(\langle a^\dagger a\rangle+\langle a a^\dagger\rangle)-\langle a^\dagger\rangle \langle a\rangle & \langle a^2\rangle -\langle a\rangle^2  \\
 \langle a^{\dagger 2}\rangle-\langle a^\dagger\rangle^2 & \frac{1}{2}(\langle a^\dagger a\rangle+\langle a a^\dagger\rangle)-\langle a^\dagger\rangle \langle a\rangle \\
\end{bmatrix},
\end{align}
is the covariance matrix. For Gaussian states the Wigner function is always positive, then, we can define the following quantities
\begin{align}
    S_{\rm W}&=-\int d^2\alpha  W(\vec{\alpha})\ln W(\vec{\alpha})=1+\ln\pi+\frac{1}{2}\ln(\det\Theta),\label{wentropy}\\
    K[W_1||W_2]&=\int d^2\alpha W_1(\vec{\alpha})\ln \frac{W_1(\vec{\alpha})}{W_2(\vec{\alpha})}\nonumber\\
   &=-1+\frac{1}{2}\left[(\vec{v}_1-\vec{v}_2)^\dagger\Theta^{-1}_2(\vec{v}_1-\vec{v}_2)+{\rm Tr}\{\Theta_2^{-1}\Theta_1\}+\ln\frac{\det\Theta_2|}{\det\Theta_1}\right],\label{wrelat}
\end{align}
which are, respectively, the Wigner entropy, equivalent to the Rényi-2 entropy, and the Wigner relative entropy between any two Gaussian states represented by $W_i(\vec{\alpha})$.

A Taylor expansion of the Wigner relative entropy $K[W_{\theta+d\theta}||W_{d\theta}]$ up to second order in $d\theta$ yields
\begin{align}
K[W_{\theta+d\theta}||W_{d\theta}]=\int d^2\alpha W_{\theta+d\theta}(\vec{\alpha})\ln \frac{W_{\theta+d\theta}(\vec{\alpha})}{W_\theta(\vec{\alpha})}\approx  \frac{1}{2}I_{\rm W}(\theta)d\theta^2  \label{wfisheri}
\end{align}
where $I_{\rm W}(\theta)$ is the Wigner-Fisher information 
\begin{align}
I_{\rm W}(\theta)=\int d^2\alpha W_{\theta}(\vec{\alpha})\left[\frac{d}{d\theta}\ln W_{\theta}(\vec{\alpha})\right]^{2}.
\end{align}
To obtain the WFI in terms of the mean vector and covariance matrix, we expand in Taylor the last equality of Eq. \eqref{wrelat}, which reads
\begin{align}
K[W_{\theta+d\theta}||W_\theta]=-1+\frac{1}{2}\left[(\vec{v}_{\theta+d\theta}-\vec{v}_\theta)^\dagger\Theta^{-1}_\theta(\vec{v}_{\theta+d\theta}-\vec{v}_\theta)+{\rm Tr}\{\Theta_\theta^{-1}\Theta_{\theta+d\theta}\}+\ln\frac{\det\Theta_\theta}{\det\Theta_{\theta+d\theta}}\right].\label{wrelat2}
\end{align}
We now expand $\vec{v}_{\theta+d\theta}$, $\Theta_{\theta+d\theta}$ and $\ln(\det\Theta_{\theta+d\theta})$ up to second order in $d\theta$
\begin{align}
    &\vec{v}_{\theta+d\theta}\approx \vec{v}_{\theta}+\left(\frac{d\vec{v}_{\theta}}{d\theta}\right)d \theta+\frac{1}{2}\left(\frac{d^2\vec{v}_{\theta}}{d\theta^2}\right)d \theta^2,\\
    &\Theta_{\theta+d\theta}\approx \Theta_{\theta}+\left(\frac{d\Theta_{\theta}}{d\theta}\right)d \theta+\frac{1}{2}\left(\frac{d^2\Theta_{\theta}}{d\theta^2}\right)d \theta^2,\\
    &\ln(\det\Theta_{\theta+d\theta})\approx \ln(\det\Theta_{\theta})+\frac{d}{d\theta}[\ln(\det\Theta_{\theta})]d \theta+\frac{1}{2}\frac{d^2}{d\theta^2}[\ln(\det\Theta_{\theta})]d \theta^2.
\end{align}
Using the above results, we now compute the expansion of each term on Eq. \eqref{wrelat2} up to second order in $d\theta$. From now on, we also adopt the over dot notation for the $\theta$-derivative, i.e., $\dot{f}=\frac{df}{d\theta}$
\begin{enumerate}
    \item Mean vector term:
    \begin{align}
    (\vec{v}_{\theta+d\theta}-\vec{v}_\theta)^\dagger\Theta^{-1}_\theta(\vec{v}_{\theta+d\theta}-\vec{v}_\theta)&\approx  \left[\left(\frac{d\vec{v}_{\theta}}{d\theta}\right)d \theta+\frac{1}{2}\left(\frac{d^2\vec{v}_{\theta}}{d\theta^2}\right)d \theta^2\right]^\dagger\Theta^{-1}_\theta\left[\left(\frac{d\vec{v}_{\theta}}{d\theta}\right)d \theta+\frac{1}{2}\left(\frac{d^2\vec{v}_{\theta}}{d\theta^2}\right)d \theta^2\right]\nonumber\\
&\approx\dot{\vec{v}}_\theta^\dagger\Theta^{-1}_\theta\dot{\vec{v}}_\theta d\theta^2.
    \end{align}
\item The trace term:
\begin{align}
    {\rm Tr}\{\Theta_\theta^{-1}\Theta_{\theta+d\theta}\}&\approx{\rm Tr}\left\{\Theta_\theta^{-1}\left[\Theta_{\theta}+\left(\frac{d\Theta_{\theta}}{d\theta}\right)d \theta+\frac{1}{2}\left(\frac{d^2\Theta_{\theta}}{d\theta^2}\right)d \theta^2\right]\right\}\nonumber\\
    &=2+{\rm Tr}\{\Theta^{-1}_\theta\dot{\Theta}_\theta\}d\theta+\frac{1}{2}{\rm Tr}\{\Theta^{-1}_\theta\ddot{\Theta}_\theta\}d\theta^2.
\end{align}
\item For the logarithm term, we use the following identities \cite{Petersen2012MatrixCookbook}
\begin{align}
&\frac{d}{d\theta}(\det\Theta_\theta)=\det\Theta_\theta{\rm Tr}\{\Theta_\theta^{-1}\dot{\Theta}_\theta\}\\
&\frac{d}{d\theta}\Theta_\theta^{-1}=-\Theta^{-1}_\theta\dot{\Theta}_\theta\Theta_\theta^{-1}\\
&\implies \frac{d}{d\theta}\ln[(\det\Theta_\theta)]={\rm Tr}\{\Theta_\theta^{-1}\dot{\Theta}_\theta\}\\
&\implies \frac{d^2}{d\theta^2}\ln[(\det\Theta_\theta)]={\rm Tr}\{\Theta_\theta^{-1}\ddot{\Theta}_\theta-\Theta_\theta^{-1}\dot{\Theta}_\theta\Theta_\theta^{-1}\dot{\Theta}_\theta\}.
\end{align}
Then
\begin{align}
\ln[\det(\Theta_{\theta+d\theta})]\approx \det\Theta_{\theta}+{\rm Tr}\{\Theta_\theta^{-1}\dot{\Theta}_\theta\}d \theta+\frac{1}{2}{\rm Tr}\{\Theta_\theta^{-1}\ddot{\Theta}_\theta-\Theta_\theta^{-1}\dot{\Theta}_\theta\Theta_\theta^{-1}\dot{\Theta}_\theta\}d \theta^2,
\end{align}
and finally
\begin{align}
\ln\left(\frac{\det\Theta_\theta}{\det\Theta_{\theta+d\theta}}\right)\approx -{\rm Tr}\{\Theta_\theta^{-1}\dot{\Theta}_\theta\}d \theta-\frac{1}{2}{\rm Tr}\{\Theta_\theta^{-1}\ddot{\Theta}_\theta\}d\theta^2+{\rm Tr}\{(\Theta_\theta^{-1}\dot{\Theta}_\theta)^2\}d \theta^2.
\end{align}
\end{enumerate}
By putting all the above results  together, we have
\begin{align}
    K[W_{\theta+d\theta}|W_\theta]\approx\frac{1}{2}\left[\dot{\vec{v}}_\theta^\dagger\Theta^{-1}_\theta\dot{\vec{v}}_\theta +{\rm Tr}\{(\Theta_\theta^{-1}\dot{\Theta}_\theta)^2\}\right]d\theta^2\label{wrelat3}.
\end{align}
Finally, by comparing \eqref{wfisheri} and \eqref{wrelat3}, we can write an analytical form for the Wigner-Fisher information in terms of $\vec{v}_\theta$ and $\Theta_\theta$ as
\begin{align}
I_{\rm W}(\theta)=\dot{\vec{v}}_{\theta}^{\dagger}\Theta_{\theta}^{-1}\dot{\vec{v}}_{\theta}+\text{Tr}\{(\Theta_{\theta}^{-1}\dot{\Theta}_{\theta})^{2}\},\label{wfiapa}
\end{align}
which is Eq. \eqref{WFI} in the main text.
\section{Passive and ergotropic contribution to the WFI}
\label{AppB}

In this section, we show that we can split the WFI into passive and ergotropic contributions, $I_{\rm W}(\theta)=I_{\pi}(\theta)+I_{\mathcal{E}}(\theta)$. First, from Eq. \eqref{wfiapa} it is clear that $I_\pi(\theta)$ must come from the term $\text{Tr}\{(\Theta_{\theta}^{-1}\dot{\Theta}_{\theta})^{2}\}$ since this is the only term that is non zero when the state is thermal. This term is also responsible for the contribution $I_{\rm s}(\theta)$ as it depends only on the covariance matrix and that $\dot{\vec{v}}_{\theta}^{\dagger}\Theta_{\theta}^{-1}\dot{\vec{v}}_\theta=0$ when no displacement is present ($\mu=0$). Using the identity
\begin{align}
\text{Tr}\{A^2\}=\text{Tr}\{A\}^2-2 \det A,\label{id}
\end{align}
which is valid for any matrix $A$ with dim-2x2, and that
\begin{align}
    \dot{S}_{\rm W}(\theta)=\frac{1}{2}\frac{d}{d\theta}[\ln(\det\Theta_\theta)],
\end{align}
we can write
\begin{align}
\text{Tr}\{(\Theta_{\theta}^{-1}\dot{\Theta}_{\theta})^{2}\}=2\dot{S}_{\rm W}^2(\theta)+\frac{1}{2}\text{Tr}\{\Theta^{-1}_\theta \dot{\Theta}_\theta\}^2-2\frac{\det \dot{\Theta}_\theta}{\det\Theta_\theta}.  \label{sep1}  
\end{align}
For any Gaussian state we have that $\det\Theta=\det\Theta_\pi$, where $\Theta_\pi$ is the covariance matrix for the associated passive state, which is thermal \cite{Medina2025}. Then,  the energy of a passive state can be written as $E_\pi(\theta)=\omega\sqrt{\det\Theta_\theta}$. This implies that the first term in Eq. \eqref{sep1} reads $2 \dot{S}^2_{\rm W}(\theta)=2[\dot{E}_\pi(\theta)/E_\pi(\theta)]^2$ - it depends only on the passive state energy. Now, when no squeezing is present, the covariance matrix is the same as for a thermal states, i.e., $\Theta_\theta^{\rm th}=[\bar{n}_{\rm th}(\theta)+1/2]\mathbb{I}$, where $\bar{n}_{\rm th}(t)=[\bar{n}_\pi(0)-\bar{n}_{\rm eq}]e^{-\Gamma t}+\bar{n}_{\rm eq}$. So, without squeezing, the second term in Eq. \eqref{sep1} is given by
\begin{align}
\frac{1}{2}\text{Tr}\{\Theta^{{{\rm th}}^{-1}}_\theta \dot{\Theta}_\theta^{\rm th}\}^2-2\frac{\det \dot{\Theta}^{\rm th}_\theta}{\det\Theta^{\rm th}_\theta}=\frac{\dot{\bar{n}}_{\rm th}^2(\theta)}{2[\bar{n}_{\rm th}(\theta)+1/2]^2}\text{Tr}\{\mathbb{I}\}^2   -2\frac{\dot{\bar{n}}_{\rm th}^2(\theta)}{[\bar{n}_{\rm th}(\theta)+1/2]^2}=0,
\end{align}
where $\mathbb{I}$ is the identity operator. Then, we can write $\text{Tr}\{(\Theta_{\theta}^{-1}\dot{\Theta}_{\theta})^{2}\}=I_\pi(\theta)+I_{\rm s}(\theta)$, where
\begin{align}
    I_\pi(\theta)&=2\dot{S}_{\rm W}^2(\theta),\\
    I_{\rm s}(\theta)&=\frac{1}{2}\text{Tr}\{\Theta^{-1}_\theta \dot{\Theta}_\theta\}^2-2\frac{\det \dot{\Theta}_\theta}{\det\Theta_\theta},
\end{align}
or, by using identity \eqref{id}, $I_{\rm s}(\theta)$ can be cast in a more symmetrical shape as
\begin{align}
I_{\rm s}(\theta)=\text{Tr}\{(\Theta_{\theta}^{-1}\dot{\Theta}_{\theta})^{2}\}-\frac{1}{2}\text{Tr}\{\Theta_{\theta}^{-1}\dot{\Theta}_{\theta}\}^2. \label{is}
\end{align}

Similarly, it is also clear that the displacement and mixed contributions, $I_{\rm d}(\theta)$ and $I_{\rm ds}(\theta)$, must come from the term $\dot{\vec{v}}_{\theta}^{\dagger}\Theta_{\theta}^{-1}\dot{\vec{v}}_{\theta}$ in Eq. \eqref{wfiapa}. For a displaced thermal state, we have only the $I_{\rm d}(\theta)$. As $\Theta_\theta$ is not affected by the displacement operation, $D(\mu)=e^{\mu\hat{a}^\dagger-\mu^*\hat{a}}$, $\Theta_\theta=\Theta_\theta^{\rm th}$, and then $\Theta_\theta^{-1}=[\sqrt{\det\Theta_\theta}\mathbb{I}]^{-1}$. The displacement contribution, is given by
\begin{align}
    I_{\rm d}(\theta)=\frac{|\dot{\vec{v}}_{\theta}|^2}{\sqrt{\det\Theta_\theta}}.
\end{align}
For a displaced and squeezed thermal state, we are going to have the contribution $I_{\rm ds}(\theta)$, which depends on the both squeezing and displacement parameters. This contribution can be obtained by subtracting $I_{\rm d}(\theta)$ from $\dot{\vec{v}}_{\theta}^{\dagger}\Theta_{\theta}^{-1}\dot{\vec{v}}_{\theta}$, which reads
\begin{align}
    I_{\rm ds}(\theta)&=\dot{\vec{v}}_{\theta}^{\dagger}\Theta_{\theta}^{-1}\dot{\vec{v}}_{\theta}-I_{\rm d}(\theta)\nonumber\\
    &=\frac{\left[({\rm Tr\{\Theta_\theta\}}-\sqrt{\det\Theta_\theta})|\dot{\vec{v}}_{\theta}|^2-\dot{\vec{v}}_{\theta}^{\dagger}\Theta_{\theta}\dot{\vec{v}}_{\theta}\right]}{\det\Theta_\theta},
\end{align}
where we used the identity
\begin{align}
    A^{-1}=\frac{1}{\det A}\left[\text{Tr}\{A\}\mathbb{I}-A\right].
\end{align}

Finally, we can explicitly compute each contribution for a general Gaussian state, $\rho_{\rm G}=D(\mu)S(z)\rho_{\pi}S^\dagger(z)D^\dagger(\mu)$. From now on we explicitly use $\theta\rightarrow t$. The mean vector and covariance matrix of this state is given by
\begin{align}
 &\vec{v}_t=(\mu_t,\mu_t^*) \ {\rm with} \ \mu_t=|\mu| e^{-(i\omega+\Gamma/2)t+i\frac{\phi_{\rm d}}{2}},\\
 &\Theta_t=
\left(\bar{n}_\pi(t)+\frac{1}{2}\right)\begin{bmatrix}
\cosh(2r_t) & -e^{i(\phi_{\rm s}-2\omega t)}\sinh(2r_t)  \\
-e^{-i(\phi_{\rm s}-2\omega t)}\sinh(2r_t)  & \cosh(2r_t)
\end{bmatrix},
\end{align}
where we note that the displaced parameter and the squeezing parameter do not mix. The parameters $r_t$ and $f(\beta_t)$ are obtained by solving the Master Eq. \eqref{meq} and are given by
\begin{align}
&\bar{n}_\pi(t)+\frac{1}{2}=\sqrt{\left[\bar{n}_{\rm th}(t)+\frac{1}{2}\right]^2+\bar{n}_{\rm s}e^{-2\Gamma t}(e^{\Gamma t}-1)},\\
&\cosh(2r_t)=\frac{2}{2\bar{n}_\pi(t)+1}\left[\bar{n}_{\rm th}(t)+(2\bar{n}_\pi(0)+1)e^{-\Gamma t}\sinh^2(r)\right],
\end{align}
where $\bar{n}_s=(2\bar{n}_\pi(0)+1)(2\bar{n}_{\rm eq}+1)\sinh^2(r)$. It is important to note that as $[D(\mu),S(z)]\neq0$, the state $\rho_{\bar{\rm G}}=S(z)D(\mu)\rho_{\rm th}D^\dagger(\mu)S^\dagger(z)$ is in general different from $\rho_{\rm G}$. However, as the following relation between displacement and squeezing operation is satisfied,
\begin{align}
    D(\alpha)S(z)=S(z)D(\gamma), \ \ \gamma=\alpha\cos(r)+\alpha^*e^{i\phi_{\rm s}}\sinh(r),
\end{align}
we can always represent a general Gaussian state by  $\rho_{\bar{\rm G}}$ or $ \rho_{\rm G}$ by choosing the appropriate displacement parameter. By using Eq. \eqref{is}, the squeezing contribution is given by
\begin{align}
    I_{\rm s}(t)= 8\dot{r}_t^2+8\omega \sinh^2(2r_t).
\end{align}
The contribution for the ergotropy coming from the squeezing operation is given by \cite{Medina2025} 
\begin{align}
    \mathcal{E}_{\rm s}(t)=E_\pi(t)[\cosh(2r_t)-1],
\end{align}
and we can define the ratio $R_{\rm s}(t)=\mathcal{E}_{\rm s}(t)/E_\pi(t)$. Then, using some algebra and identities for hyperbolic functions, we have that
\begin{align}
&\sinh^2(2r_t)=R_{\rm s}(t)[R_{\rm s}(t)+2],\\
&\dot{r}_t^2= \frac{\dot{R}_{\rm s}(t)}{4R_{\rm s}(t)[R_{\rm s}(t)+2]}.
\end{align}
Using the above results, we can rewrite the squeezing contribution to the WFI only in terms of $R_{\rm s}(t)$ as
\begin{align}
I_{\rm s}(t)=\frac{2\dot{R}_s^{2}(t)}{R_s(t)[2+R_s(t)]}+8\omega^2R_s(t)[2+R_s(t)],
\end{align}
which is Eq. (17) in the main text.

To obtain the contributions $I_{\rm d}(t)$ and $I_{\rm ds}(t)$ we can directly compute 
\begin{align}
\dot{\vec{v}}_\theta^\dagger\Theta^{-1}_\theta\dot{\vec{v}}_\theta&=\frac{4}{2\bar{n}_\pi(t)+1}\left[|\dot{\mu}_t|^2\cosh(2r_t)+\sinh(2r_t)\Re[\dot{\mu}_t^2e^{-i(\phi_{\rm s}-2\omega t)}]\right].
\end{align}
Now, we need to proceed with some algebraic simplifications. Let us first work with the second term inside the brackets. First, we compute $\dot{\mu}^2_t$:
\begin{align}
\dot{\mu}^2_t=\left[\frac{\Gamma}{2}+i\omega\right]^2|\mu_t|^2e^{i(\phi_{\rm d}-2\omega t)}.
\end{align}
Then
\begin{align}
    \Re[\dot{\mu}^2_t e^{-i(\phi_{\rm s}-2\omega t)}]&=|\mu_t|^2\Re\left[\left(\frac{\Gamma}{2}+i\omega\right)^2e^{i\Delta}\right],\nonumber
\end{align}
where we defined the difference between the displacement and squeezing phases  as $\Delta=\phi_{\rm d}-\phi_{\rm s}$. Then
\begin{align}
\dot{\vec{v}}_\theta^\dagger\Theta^{-1}_\theta\dot{\vec{v}}_\theta=\frac{4}{2\bar{n}_\pi(t)+1}\left[|\dot{\mu}_t|^2\cosh(2r_t)+|\mu_t|^2\sinh(2r_t)\Re\left[\left(\frac{\Gamma}{2}+i\omega\right)^2e^{i\Delta}\right]\right].\label{crossterm}
\end{align}
Let us simplify the phase term
\begin{align}
\Re\left[\left(\frac{\Gamma}{2}+i\omega\right)^2 e^{i\Delta}\right]&=\Re\left[\left(\frac{\Gamma}{2}+i\omega\right)^2 \right]\cos\Delta+\Re\left[i\left(\frac{\Gamma}{2}+i\omega\right)^2 \right]\sin\Delta\nonumber\\
&=A \cos\Delta+B\sin\Delta\nonumber\\
&=\sqrt{A^2+B^2}\cos(\Delta-\varphi),
\end{align}
where we defined $A=\Re\left[\left(\frac{\Gamma}{2}+i\omega\right)^2 \right]$, $B=\Re\left[i\left(\frac{\Gamma}{2}+i\omega\right)^2\right]$ and then $\tan\varphi= A/B$. Explicitly, we have
\begin{align}
    &\sqrt{A^2+B^2}=\frac{\Gamma^2}{4}+\omega^2,\\
    &\tan\varphi=\frac{\Gamma}{4\omega}-\frac{\omega}{\Gamma}.
\end{align}
Substituting this in Eq. \eqref{crossterm}, we have
\begin{align}
\dot{\vec{v}}_\theta^\dagger\Theta^{-1}_\theta\dot{\vec{v}}_\theta&=\frac{4}{2\bar{n}_\pi(t)+1}\left[|\dot{\mu}_t|^2\cosh(2r_t)+|\mu_t|^2\left(\frac{\Gamma^2}{4}+\omega^2\right)\sinh(2r_t)\cos(\Delta-\varphi)\right].\label{idids}
\end{align}
In Eq. \eqref{idids}, we identify
\begin{align}  |\dot{\mu}_t|^2=|\mu_t|^2\left(\frac{\Gamma^2}{4}+\omega^2\right),
\end{align}
so, as consequence, we have that
\begin{align}
\dot{\vec{v}}_\theta^\dagger\Theta^{-1}_\theta\dot{\vec{v}}_\theta&=\frac{4|\dot{\mu}_t|^2}{2\bar{n}_\pi(t)+1}\left[\cosh(2r_t)+\sinh(2r_t)\cos(\Delta-\varphi)\right]\nonumber\\
&=I_{\rm d}(t)\left[(R_{{\rm s}}(t)+1)+\sqrt{R_{{\rm s}}(t)[R_{{\rm s}}(t)+2]}\cos(\Delta-\varphi)\right]\nonumber\\
&=I_{\rm d}(t)+I_{\rm ds}(t)
\end{align}
where
\begin{align}
&I_{\rm d}(t) = \frac{|\dot{\vec{v}}_t|^2}{\sqrt{\det \Theta_t}}=\frac{4|\dot{\mu}_t|^2}{2\bar{n}_\pi(t)+1}=2\left(\frac{\Gamma^2}{4}+\omega^2\right)R_{\rm d}(t),\\
&I_{\rm ds}(t)=I_{\rm d}(t)\left[R_{{\rm s}}(t)+\sqrt{R_{{\rm s}}(t)[R_{{\rm s}}(t)+2]}\cos(\Delta-\varphi)\right],
\end{align}
with $R_{\rm d}(t)=\mathcal{E}_{\rm d}(t)/E_\pi(t)$ and displacement ergotropy $\mathcal{E}_{\rm d}(t)=\omega |\mu_t|^2$ \cite{Medina2025}.
\section{Connection between the Wigner entropy production and the ergotropic loss rate}
\label{AppC}

To establish this connection, we first recall that the mean energy for a Gaussian state is given by \cite{Singh2019}
\begin{align}
    E(t)=\frac{\omega}{2}[|\vec{v}_t|^2+{\rm Tr}\{\Theta_t\}].
\end{align}
Now, from Eq. \eqref{wrelat}, we can compute the entropy production for a Gaussian state, which is defined as \cite{Santos2017}
\begin{align}
    \Pi_{\rm W}(t)&=-\frac{d}{dt}K[W||W_{\rm eq}]\nonumber\\
    &=-\frac{1}{E_{\rm eq}}\frac{d}{dt}\left[\frac{\omega}{2}(|\vec{v}_t|^2+{\rm Tr}\{\Theta_t\})\right]-\frac{1}{2}\frac{d}{dt}[\ln (\det\Theta_{\rm eq})-\ln(\det\Theta_t)]\nonumber\\
    &=-\frac{\dot{E}(t)}{E_{\rm eq}}+\dot{S}_{\rm W}(t),\label{entprod}
\end{align}
where $E_{\rm eq}=\omega(\bar{n}_{\rm eq}+1/2)$ is the mean energy of the equilibrium state. Here, we remark that the first term in Eq. \eqref{entprod} is the entropy flux rate, i.e., $\Phi=-\frac{\dot{E}(t)}{E_{\rm eq}}$. For Gaussian states, every passive state is a thermal state with temperature determined from $\sqrt{\det\Theta_\pi}=\sqrt{\det\Theta_t}=\bar{n}_\pi(t)+1/2$. We can then define the entropy production for the passive state as
\begin{align}
 \Pi_{\pi}(t)&=-\frac{d}{dt}K[W_\pi||W_{\rm eq}]\nonumber\\
 &=-\frac{\dot{E}_\pi(t)}{E_{\rm eq}}+\dot{S}_{\pi}(t)\label{entprodpass}
\end{align}
Now, by subtracting \eqref{entprod} from \eqref{entprodpass}, recalling that $S_{\rm W}(t)=S_{\pi}(t)$ and using that the ergotropy is defined by $\mathcal{E}(t)=\omega(\bar{n}_\pi(t)+1/2)K[W||W_\pi]=E(t)-E_\pi(t)$, we have that
\begin{align}
    \dot{\mathcal{E}}(t)=-E_{\rm eq}[\Pi_{\rm W}(t)-\Pi_{\pi}(t)],
\end{align}
or, equivalently,
\begin{align}
    \frac{d}{dt}E_\pi(t)K[W||W_\pi]=-\frac{d}{dt}\left[K[W||W_{\rm eq}]-K[W_\pi||W_{\rm eq}]\right].
\end{align}
\end{document}